# Ultrafast absorptive and refractive nonlinearities in multi-walled carbon nanotube film

H. I. Elim, W. Ji*, G. H. Ma, K. Y. Lim, C. H. Sow, and C. H. A. Huan

Department of Physics
National University of Singapore
2 Science Drive 3, Singapore 117542
Republic of Singapore

## Abstract

By using femtosecond laser pulses at a wavelength range from 720 to 780 nm, we have observed absorptive and refractive nonlinearities in a film of multi-walled carbon nanotubes grown mainly along the direction perpendicular to the surface of quartz substrate. The Z-scans show that both absorptive and refractive nonlinearities are of negative and cubic nature in the laser irradiance range from a few to 300 GW/cm$^2$. The magnitude of the third-order nonlinear susceptibility, $\chi^{(3)}$, is of an order of $10^{-11}$ esu. The degenerate pump-probe measurement reveals a relaxation time of ~ 2 ps.

---

* Email address: phyjiwei@nus.edu.sg

In the last decade, fundamental physical properties of carbon nanotubes have attracted much attention because of their tremendous promise in a wide range of technological applications [1-11]. In particular, ultrafast nonlinear optical responses of single-wall carbon nanotubes (SWCNTs) in suspensions and in films have been investigated intensively in recent years [12-17]. Transient photobleaching has been observed with femtosecond laser pulses at photon energies of 0.8 ~ 1.1 eV (wavelength $\lambda$ = 1.1 to 1.5 μm), resonant with the lowest inter-band transitions of semiconducting SWCNTs. The imaginary part of the third-order nonlinear susceptibility, $\mathrm{Im}\chi^{(3)}$, has been determined to be as large as $10^{-7}$ esu [16]. Saturable absorption has also been detected at photon energies of ~ 1.47 eV ($\lambda$ ~ 843 nm), near the second lowest inter-band transitions of semiconducting electronic structure in SWCNTs [14]. In addition, transient photoinduced absorption has been reported under off-resonant conditions. The resonant saturable absorption is identified as a band-filling mechanism, while the off-resonant photoinduced absorption is attributed to a global redshift of the $\pi$-plasmon resonance [14].

Note that all the above-mentioned investigations focus on the imaginary part of $\chi^{(3)}$ and its picosecond (or subpicosecond) relaxation. It is known that the real part of the third-order nonlinear susceptibility, $\mathrm{Re}\chi^{(3)}$, is another key parameter if carbon nanotubes are to be used as ultrafast switching materials. It is generally accepted that multi-walled carbon nanotubes (MWCNTs) behave like metals. We expect that the $\pi$-electrons should play an important role in the electronic structures and optical properties of MWCNTs. In SWCNTs, the $\pi$-plasmon resonance is located at ~5 eV and spans the entire visible spectral region with a tail extending to near-infrared wavelengths [14]. The $\pi$-plasmon is a collective excitation of the $\pi$ electrons. In a bulk medium, this resonance is not coupled to light. But, in the case of nanoscale objects, it changes into a surface plasmon. In metallic nanoparticles like silver particles, it gives rise to large values of $\mathrm{Im}\chi^{(3)}$ and $\mathrm{Re}\chi^{(3)}$ [18]. We believe that similar nonlinear optical processes should occur in MWCNTs. Here we report the observation of negative absorptive and refractive nonlinearities in a film of MWCNTs under off-resonant conditions. These cubic nonlinearities also exhibit ultrafast relaxation. All these findings strongly suggest the potential of MWCNT films for all-optical switching.

The MWCNT film grown on quartz substrate was prepared by a method of plasma-enhanced chemical vapor deposition. The details of the preparation were reported elsewhere [19,20]. Figure 1(a) shows a high magnification scanning electron microscopy (SEM) image on the side view of the film and substrate. It shows that the carbon nanotubes, like bushes, are grown mainly in the direction perpendicular to the surface of the substrate, although some of the tubes are entangled with one another. The size distribution of the as-grown sample is displayed in Fig. 1(b), showing an average diameter of ~ 40 nm and an average length (or film thickness) of 1.3 μm. The linear and nonlinear optical properties of the MWCNT film were examined as the laser light propagates in the axis perpendicular to the quartz substrate (or normal incidence on the film) at room temperature. Figure 1(c) displays the absorption spectrum of the MWCNT film at photon energies between 1 and 6 eV. It clearly demonstrates that the optical property of the MWCNT film is dominated by an absorption resonance peaked at 5.4 eV. It is assigned as the π-plasmon resonance, which is about 0.4 eV blue shifted in comparison to that of SWCNTs measured by Lauret et al. [14]. The blue shift is anticipated since there are more π-electrons in MWCNTs than in SWCNTs.

To minimize average power and reduce accumulative thermal effects, we employed 100-fs laser pulses at 1 kHz repetition rate. The laser pulses were generated by a mode-locked Ti: Sapphire laser (Quantronix, IMRA), which seeded a Ti: Sapphire regenerative amplifier (Quantronix, Titan). The wavelengths were tunable as the laser pulses passed through an optical parametric amplifier (Quantronix, TOPAS). The laser pulses were focused onto the MWCNT film with a minimum beam waist of ~ 42 μm. The incident and transmitted laser powers were monitored as the MWCNT film was moved (or Z-scanned) along the propagation direction of the laser pulses. Figure 2 displays typical open- and closed-aperture Z-scans, showing negative signs for the absorptive and refractive nonlinearities. We assume that the observed absorptive and refractive nonlinearities can be described by $\Delta\alpha = \alpha_2 I$ and $\Delta n = n_2 I$ , respectively; where $\alpha_2$ is the nonlinear absorption coefficient, $n_2$ is the nonlinear refractive index, and I is the light irradiance. Both the $\alpha_2$ and $n_2$ values are extracted from the best fittings between the data and the Z-scan theory [21]. We plot the extracted $\alpha_2$ and $n_2$ values as a function of the maximum irradiance for each Z-scan. As shown in Fig. 3, it is evident that the $\alpha_2$ and $n_2$ parameters are independent of the irradiance, indicating a pure third-order nonlinear optical process in the irradiance regime up to 300 GW/cm$^2$. No observation of saturation in $\alpha_2$ and $n_2$ clearly demonstrates that the underlying mechanism in MWCNTs is different from resonant saturable absorption in SWCNTs, in which the saturation in the inter-band transitions of semiconducting electronic structures plays an important role [14,17]. Within our experimental errors, Fig. 2 shows little change in the magnitudes of the measured $\alpha_2$ and $n_2$, as the photon energies are tuned from 1.59 to 1.72 eV (780 to 720 nm), indicating a nonresonant nature of our observed nonlinearities. This is consistent with the absorption spectrum in Fig. 1(c), which shows that the probed wavelengths are in the tail of the $\pi$-plasmon resonance.

At 780 nm, we determine that $\alpha_2$ = -29 cm/GW (or Im$\chi^{(3)}$ = - 1.6 x $10^{-11}$ esu); and $n_2$ = -3 x $10^{-4}$ cm$^2$/GW (or Re $\chi^{(3)}$ = -1.7 x $10^{-11}$ esu). Hence $|\chi^{(3)}| = [(\mathrm{Im}\,\chi^{(3)})^2 + (\mathrm{Re}\,\chi^{(3)})^2]^{1/2}$ = 2.2 x $10^{-11}$ esu for the MWCNT film. By comparison, a value of 8.3 x $10^{-13}$ esu at 800 nm was observed for $|\chi^{(3)}|$ in carbon nanoparticles embedded in sol-gel $SiO_2$ glass [22]. For a SWCNT solution at a concentration of 0.33 mg/mL, it was reported to be 4 x $10^{-13}$ esu at 820 nm [12]. Our larger value of $|\chi^{(3)}|$ is attributed to the high concentration of MWCNTs packed in a 1.3 $\mu$m-thick layer. In our previously reported closed-aperture Z-scans on solutions of MWCNT/polymer composites [11], a positive sign of $n_2$ was detected in association with thermally originated nonlinear scattering for optical limiting. The negative sign of the measured $n_2$ (or self-defocusing) implies that thermal effects are insignificant in the MWCNT film. In most of semiconductors, ultrafast off-resonant nonlinearities are dominated by two-photon absorption and bound electronic optical Kerr effect (or self-focusing effect). These nonlinear processes result in positive signs for the $\alpha_2$ and $n_2$ nonlinear parameters; and lead to laser-induced damage at high irradiances. On the contrary, the observed nonlinearities in the MWCNT film are negative and, therefore, are desirable in optical applications.

To evaluate the relaxation time and to gain an insight of the underlying mechanism for the observed cubic nonlinearities, we conducted a degenerate pump-probe experiment at 780 nm with the use of 250-fs laser pulses from a mode-locked Ti:Sapphire laser (Spectro-Physics, Tsunami). Figure 4 illustrates the transient transmission signals ($\Delta T/T$) as a function of the delay time for three pump irradiances. With the film excited by the pump pulses, an optically induced transparency occurs. As the pump irradiance increases, the signal increases. All these findings are consistent with our Z-scan measurements. The transient signals clearly shows there are two components. By using a two-exponential component model, the best fits (solid lines in Fig. 4) produce $\tau_1 = \sim 250$ fs and $\tau_2 = \sim 2$ ps. We believe that $\tau_1$ is the autocorrelation of the laser pulses used. The $\tau_2$ component is the recovery time of the excited $\pi$-electrons in the MWCNT film. Its time scale is comparable to the findings for carbon nanoparticles [22] and SWCNTs [12-17]. The nature of this relaxation is unclear at present. The relaxation through charge transfer via coherent tunneling processes to neighboring metallic tubes is possible [14] and such a relaxation has been observed in $C_{60}$ fullerene films [23]. Further studies far beyond the scope of this investigation are required to unambiguously identify the nature of the relaxation in the MWCNT film.

In summary, we have presented femtosecond Z-scan and pump-probe measurements on a 1.3-μm-thick MWCNT film, in which the nanotubes are aligned mainly in the propagation direction of the laser pulses. These measurements show the MWCNT film possesses cubic nonlinear absorption and refraction in the laser irradiance range from a few to 300 GW/cm$^2$. These nonresonant and negative nonlinearities exhibit magnitudes of $\sim 10^{-11}$ esu at 780 nm and a relaxation time of $\sim 2$ ps.

**References**

[1]. J.W. Mintmire, B.I. Dunlap, and C.T. White, Phys. Rev. Lett. **68**, 631 (1992).

[2]. M.S. Dresselhaus, Nature **358**, 195 (1992).

[3]. T.W. Ebbesen and P.M. Ajayan, Nature **358**, 220 (1992).

[4]. P.M. Ajayan, O. Stephan, C. Colliex, and D. Trauth, Science **265**, 1212 (1994).

[5]. W.A. de Heer, A. Chatelain, and D. Ugarte, Science **270**, 1179 (1995).

[6]. H. Dai, E.W. Wong, Y.Z. Lu, S. Fan, and C.M. Lieber, Nature **375**, 769 (1995).

[7]. M.M. J. Treacy, T.W. Ebbesen, and J.M. Gibson, Nature **381**, 678 (1996).

[8]. S.J. Tans, A.R. M. Verschueren, and C. Dekker, Nature **393**, 49 (1998).

[9]. X. Sun, R.Q. Yu, G.Q. Xu, T.S.A. Hor, and W. Ji, Appl. Phys. Lett. **73**, 3632 (1998).

[10]. P. Chen, X. Wu, X. Sun, J. Lin, W. Ji and K.L. Tan, Phys. Rev. Lett. **82**, 2548 (1999).

[11]. Z.X. Jin, X. Sun, G.Q. Xu, S.H. Goh, and W. Ji, Chem. Phys. Lett. **318**, 505 (2000).

[12]. S. Wang, W. Huang, H. Yang, Q. Gong, Z. Shi, X. Zhou, D. Qiang, and Z. Gu, Chem. Phys. Lett. **320**, 411 (2000).

[13]. Y.C. Chen, N.R. Raravikar, Y.P. Zhao, L.S. Schadler, P.M. Ajayan, T.M. Lu, G.C. Wang, and X.C. Zhang, Appl. Phys. Lett. **81**, 975 (2002).

[14]. J-S. Lauret, C. Voisin, G. Cassabois, C. Delalande, Ph. Roussignol, O. Jost, and L. Capes, Phys. Rev. Lett. **90**, 057404 (2003).

[15]. H. Han, S. Vijayalakshmi, A. Lan, Z. Iqbal, H. Grebel, E. Lalanne, and A.M. Johnson, Appl. Phys. Lett. **82**, 1458 (2003).

[16]. S. Tatsuura, M. Furuki, Y. Sato, I. Iwasa, M. Tian, and H. Mitsu, Adv. Mater. **15**, 534 (2003).

[17]. G.N. Ostojic, S. Zaric, J. Kono, M.S. Strano, V.C. Moore, R.H. Huage, and R.E. Smalley, submitted, (arXiv:cond-mat/0307154 v18 Jul 2003).

[18]. G. Yang, W. Wang, Y. Zhou, H. Lu, G. Yang, and Z. Chen, Appl. Phys. Lett. **81**, 3969 (2002).

[19]. Y.H. Wang, J. Lin, C.H.A. Huan, and G.S. Chen, Appl. Phys. Lett. **79**, 680 (2001).

[20]. K.Y. Lim, C.H. Sow, J. Lin, F.C. Cheong, Z.X. Shen, J.T.L. Thong, K.C. Chin, and A.T.S. Wee, Adv. Mater. **15**, 300 (2003).

[21]. M. Sheik-Bahae, A.A. Said, T. Wei, D.J. Hagan, and E.W. Van Stryland, *IEEE J. Quantum Electron* **26**, 760 (1990).

[22]. D. Li, Y. Liu, H. Yang, and S. Qian, Appl. Phys. Lett. **81**, 2088 (2002).

[23]. R.A. Cheville and N.J. Halas, Phys. Rev. B **45**, R4548 (1992)

**Figure Captions:**

Fig. 1 **(a)** High magnification SEM image, **(b)** diameter distribution with an average diameter of ~ 40 nm, and **(c)** UV-visible absorption spectrum of the MWCNT film grown on quartz substrate with an average film thickness of 1.3 μm.

Fig. 2 Z-scans of the MWCNT film on quartz measured at 720 and 780 nm by using 100-fs laser pulses focused with a minimum waist of 42 μm. The solid lines are the best-fit curves calculated by using the Z-scan theory reported in Ref. [21]. The filled and open circles in **(a)** and **(c)** are the Z-scans without aperture, while the filled and open circles in **(b)** and **(d)** are the Z-scans with aperture.

Fig. 3 Irradiance independence of **(a)** $\alpha_2$ and **(b)** $n_2$ values measured at 780 nm.

Fig. 4 Degenerate 250-fs time-resolved pump-probe measurements of the MWCNT film preformed at 780 nm with three different pump irradiances. The top and middle curves are shifted vertically for clear presentation. The solid lines are two-exponential fitting curves with $\tau_1$ = 250 fs and $\tau_2$ = ~ 2 ps.

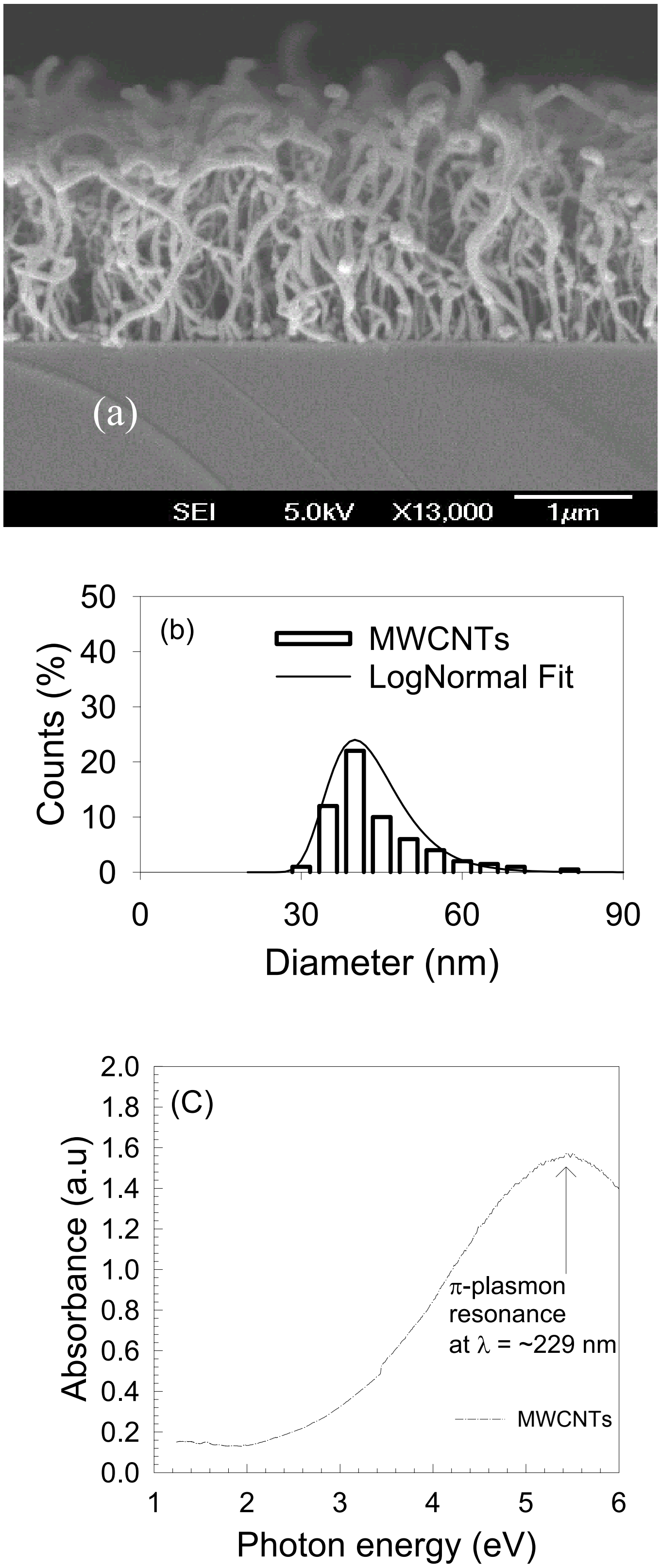

**Fig. 1.**

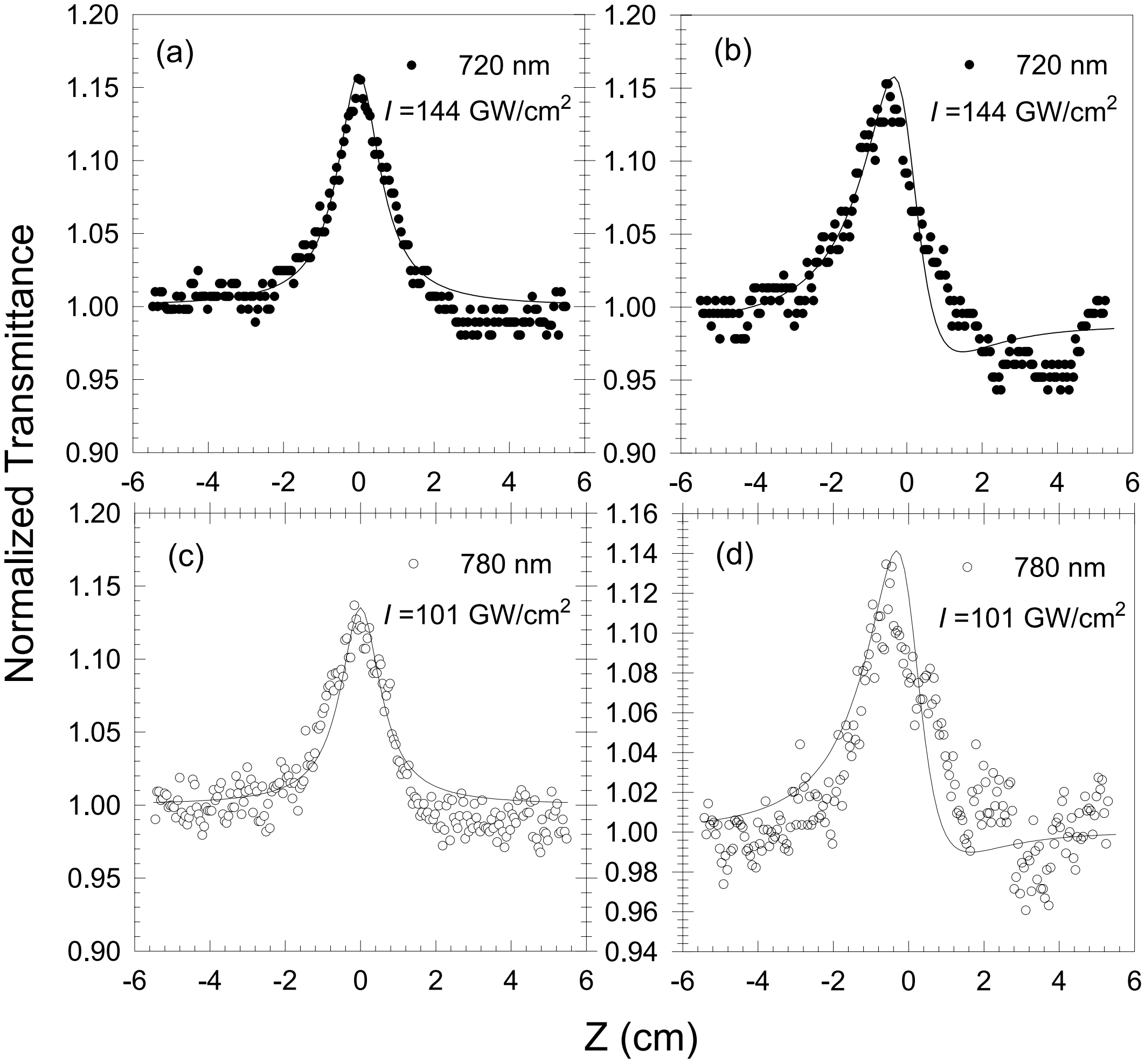

**Fig. 2.**

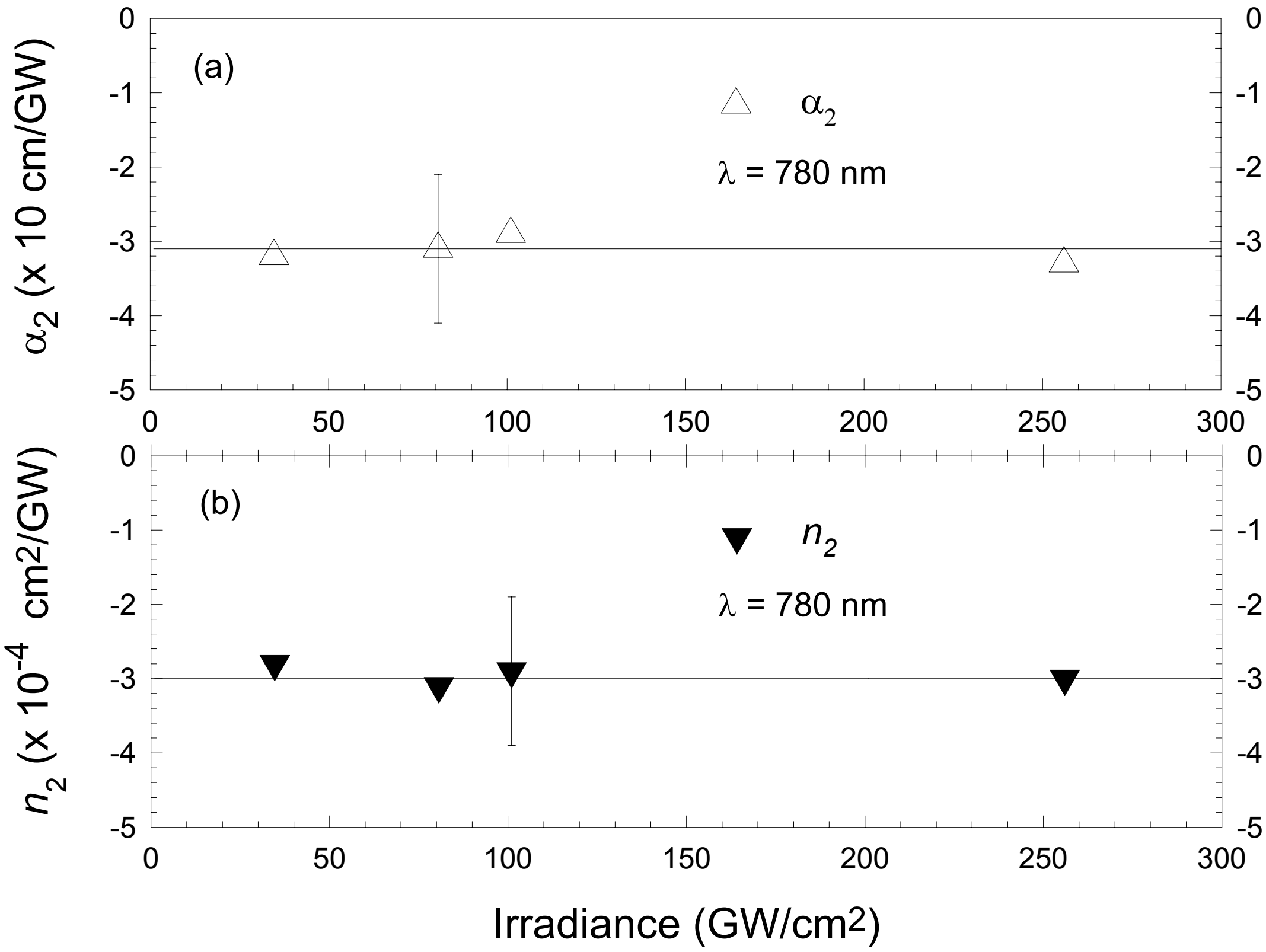

**Fig. 3.**

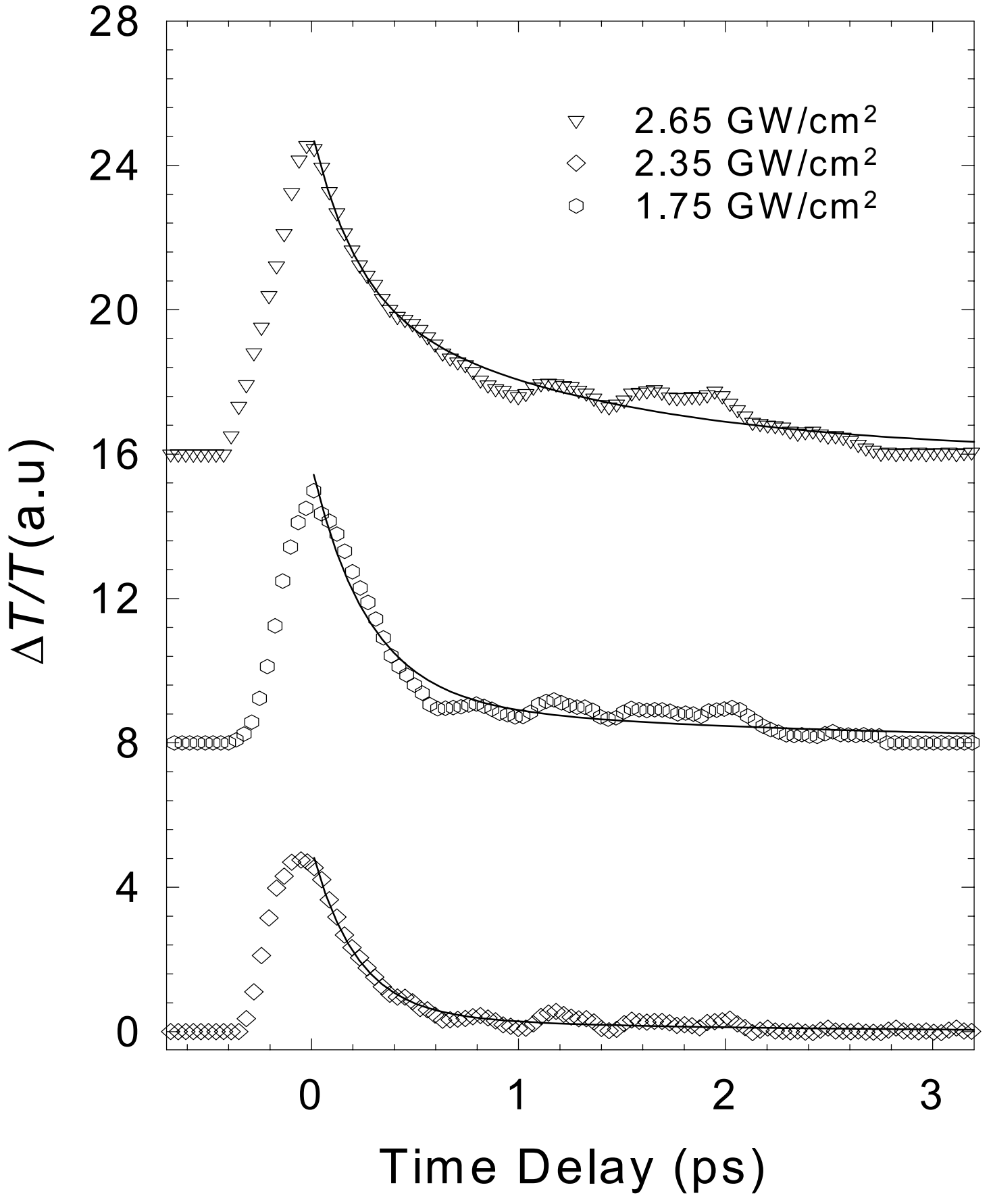

**Figure 4.**